\documentclass[conference]{IEEEtran}

\IEEEoverridecommandlockouts

\usepackage{graphicx,color}
\usepackage{amsthm,amssymb,amsmath}
\usepackage{algorithm}
\usepackage{algpseudocode}

\theoremstyle{definition}

\newtheorem{definition}{Definition}
\newtheorem{remark}{Remark}

\newcommand{\ie}{i.e.}
\newcommand{\eg}{e.g.}
\newcommand{\mc}[1]{\mathcal{#1}}
\renewcommand{\S}{Section~}

\title{On Communication Cost vs. Load Balancing in \\ Content Delivery Networks}

\author{
	\IEEEauthorblockN{M.~Jafari~Siavoshani\IEEEauthorrefmark{1}, S.~P.~Shariatpanahi\IEEEauthorrefmark{2}, H.~Ghasemi\IEEEauthorrefmark{1}, A.~Pourmiri\IEEEauthorrefmark{2}}
    \IEEEauthorblockA{\IEEEauthorrefmark{1}Department of Computer Engineering,
Sharif University of Technology, Tehran, Iran\\
mjafari@sharif.edu, hghasemi@ce.sharif.edu}
    \IEEEauthorblockA{\IEEEauthorrefmark{2}School of Computer Science,
Institute for Research in Fundamental Sciences (IPM), Tehran, Iran\\
\{pourmiri, pooya\}@ipm.ir}
}

\begin{document}
\maketitle

\begin{abstract}
It is well known that load balancing and low delivery communication cost are two critical issues in mapping requests to servers in Content Delivery Networks (CDNs).
However, the trade-off between these two performance metrics has not been yet quantitatively investigated in designing efficient request mapping schemes. In this work, we formalize this trade-off through a stochastic optimization problem.
While the solutions to the problem in the extreme cases of \emph{minimum communication cost} and \emph{optimum load balancing} can be derived in closed form, finding the general solution is hard to derive. Thus we propose three heuristic mapping schemes and compare the trade-off performance of them through extensive simulations.

Our simulation results show that at the expense of high query cost, we can achieve a good trade-off curve. Moreover, by benefiting from the \emph{power of multiple choices} phenomenon, we can achieve almost the same performance with much less query cost. Finally, we can handle requests with different delay requirements at the cost of degrading network performance.
\end{abstract}

\vspace{7pt}

\section{Introduction}\label{sec:Introduction}
Content delivery networks are becoming an indispensable architecture design of modern communication networks. As mentioned by Cisco in \cite{CiscoForcast2015}, 80\% of Internet traffic will be of multimedia nature by 2019 and by this time half of the Internet traffic will be handled by CDNs. Many practical solutions are already deployed to manage this growing demand such as Akamai \cite{Akamai_Nygren_2010}, Azure, Amazon CloudFront, and Limelight \cite{Zhang_CDN_Survey}.
In such networks, a number of servers (mainly deployed in large scale data centers) are distributed geographically which cache contents from original provider near end-users. Then, upon arrival of each user request, it will be served by an appropriate CDN server. This will reduce congestion at the original content provider.

An important design problem in CDNs is the assignment of each request to an appropriate CDN server \cite{DynamicReplicaPlacement_ICC05}. Usually an appropriate server is interpreted as the nearest one which has cached the file \cite{ChenLPCCSA05, ManfrediOR13,Chen_Sigcomm2015}. This interpretation leads to low \emph{communication cost} when delivering content to the user. However, there is another performance metric which should be considered in practice, namely, \emph{load balancing}. In other words, a practical assignment scheme should not impose a large number of requests to a single CDN server. To this end, to assign each request, the CDN mapping algorithm should consider the current load of servers. Then it should prevent assigning a request to a busy server which may result in choosing a far server from the request. Thus, these two metrics, namely, proximity and load balancing may be in contention in practical scenarios \cite{Wang2002_CDN_RequestRedirection}.

In this paper, we first formalize the trade-off between communication cost and load balancing in a CDN as a stochastic optimization problem. The solution of this problem is an assignment strategy which will result in the optimum communication cost vs. load balancing trade-off. In the extreme cases of the minimum communication cost and optimal load balancing performance the solutions can be obtained in closed form. However,
due to the complexity of obtaining the optimal solution in general, we propose three heuristic request assignment strategies which result in different trade-off curves. 

The first proposed scheme manages the trade-off between communication cost and load balancing by probabilistically switching between the two above mentioned extreme cases. The switching probability is the algorithm parameter which determines the operational point of the trade-off. In the second suggested scheme for assigning each request we define a \emph{desirability value} for each server that has cached the requested file. This desirability value takes into account both communication cost and current load of these servers. Then the request is assigned to a server with the minimum desirability value. Finally, the third scheme benefits from the notion of \emph{power of multiple choices} \cite{ABKU99} in balanced allocation literature to manage the trade-off. In this scheme for each request we find $\Delta$ servers with least communication costs for responding to that request. Then the request is assigned to the one with the minimum current load among these $\Delta$ servers.


It should be noted that although in this paper we focus on the trade-off between load balancing and communicaion cost of the proposed schemes, another important practical metric is the overhead of each scheme due to queue length queries for each assignment. Thus, in this paper, we also compare the schemes in terms of the query overhead imposed to the network.

The idea of exploiting caching servers to bring data near end-users is well known and has been used in commercial systems as well, \cite{Akamai_Nygren_2010,Zhang_CDN_Survey}. However, the technical challenges introduced by such framework is still the topic of many active research works. Load balancing in CDNs, as one of the most important issues, has been treated through various approaches. In this challenge the mapping algorithm which assigns content requests from users to CDN servers should make sure that no server becomes overloaded. Among load balancing proposals, distributed approaches, such as \cite{AdlerCMR98} \cite{ManfrediOR13}, and \cite{XiaAYL15} have attracted special attention due to their more practical nature. One promising distributed approach is \emph{randomized load balancing} benefiting from the power of multiple choices \cite{ABKU99}, \cite{Mitzenmacher01}, and \cite{Mitzen_sur01}. In this approach, for assigning each request to an appropriate server, first the current load of a few number of randomly selected servers are queried. Then, the request is assigned to the server with minimum load to balance out the load of network. 

The focus of all above works is balancing out network load without considering its interaction with the communication cost of delivering the content. In this paper, for the first time, we formalize the optimization problem which captures the fundamental trade-off between load balancing performance and communication cost.
Moreover, in contrast to previous works, our schemes are able to manage this trade-off properly.

The rest of the paper is organized as follows. In \S\ref{sec:ProbDef}, we first explain the network model and formally define our metrics. Then in \S\ref{sec:LoadBal_CommCost}, we formalize the trade-off between load balancing and communication cost through a stochastic optimization problem. Afterwards, three heuristic schemes are proposed which capture this trade-off. In \S\ref{sec:PerfEval}, we compare the performance of proposed scheme through extensive simulations. Finally, \S\ref{sec:Conclusion} concludes the paper.

\section{Problem Definition}\label{sec:ProbDef}

\begin{figure}
\begin{center}
\includegraphics[width=0.5\textwidth]{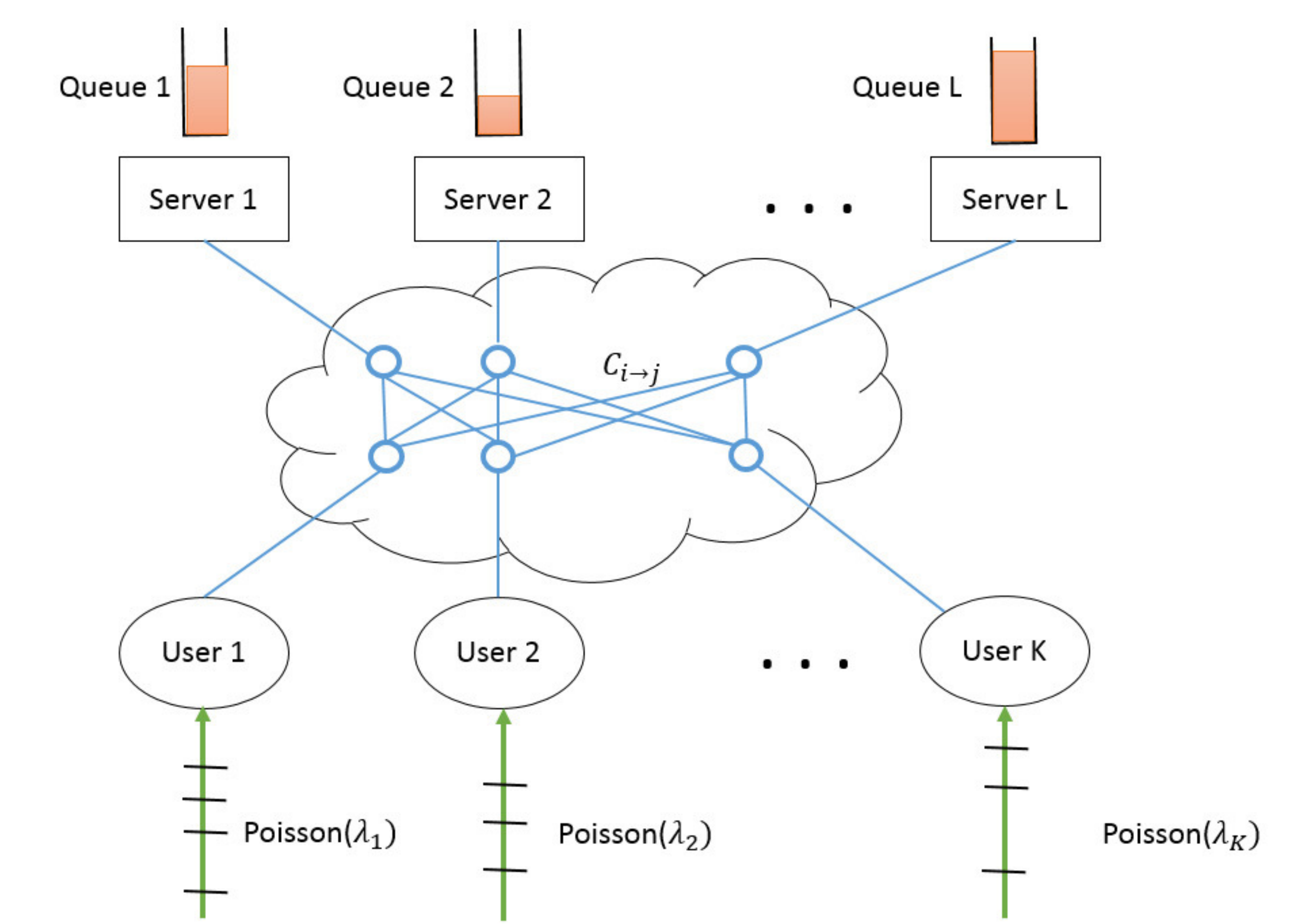}
\end{center}
\caption{A CDN with $K$ users and $L$ servers.}\label{fig:Model}
\end{figure}

We consider a content delivery network whose goal is to deliver contents from a library of $N$ files $\mc{W}=\{W_1,\dots,W_N\}$, each of $F$ bits, to the users requesting files. The network consists of three major elements (see Fig.~\ref{fig:Model}):
 \begin{itemize}
 \item
 $L$ servers each capable of storing $MF$ bits.
 \item
 $K$ users requesting files in continuous time according to $K$ Poisson point processes with parameter $\lambda_i$ for user $i\in [1:K]$. 
 \item
A communication network for transferring files from servers to users. The cost of transferring a file of $F$ bits from Server $j$ to User $i$ is $c_{i,j}$. 
For convenience, we define a cost matrix $\mathbf{C}=[c_{i,j}]_{i=1,j=1}^{i=K,j=L}$.
 \end{itemize}

In each request, the probability of requesting file $W_i$ is $p_i$. Moreover, we let $\mathcal{P}=\{p_1,\dots,p_N\}$ represent the file popularity profile.
In this paper, we assume that the file popularity profile follows the Zipf distribution  with parameter $\beta$. In Zipf distribution the request probability of the $i$-th popular file is inversely proportional to its rank as follows
\begin{equation*}
p_i=\frac{1/i^\beta}{\sum\limits_{j=1}^{K}1/j^\beta},\quad i=1,\dots,K,
\end{equation*}
which has been confirmed to be the case in many practical applications \cite{Zipf1_99,Zipf2_07}. Note that the case $\beta=0$ results in a uniform file popularity distribution.

We assume that the network operates in two phases, namely, \emph{cache content placement}, and \emph{content delivery}.  In the placement phase each server caches $M$ files. We assume $LM \geq N$ so that all files are stored in the aggregate memory of servers. The content of Server $i$'s cache is denoted by $\mc{Z}_i\subseteq \{1,\ldots,N\}$ where $| \mc{Z}_i | \le M$. In this paper we assume that in the cache content placement phase, file $W_i$ is cached with probability $p_i$, \ie, proportional cache content placement.

In the delivery phase, when a user requests file $W_i$, the request will be mapped to a server which has cached this file. Then, this request will be appended to the server's queue.  Each server's queue follows a First-In First-Out (FIFO) strategy with a given service time Probability Distribution Function (PDF) (\eg, exponential or constant).

As cache content placement phase is during network low-peak hours, the only restriction in this phase is the cache size of each server. In other words the communication cost in this phase is negligible compared to that of content delivery phase.

Suppose large enough time $T$  has passed in the delivery phase. We define

\begin{itemize}
\item
$n_i =$ the number of requests of user $i$ during this time interval. This is a random variable with $\mathbb{E}[n_i] = T\lambda_i$.
\item
$\mathbf{t}_i = (t_{i,1},\dots,t_{i,n_i}) $, where $t_{i,j}$ is the arrival time of $j$'th request of user $i$.
\item 
$\mathbf{d}_i = (d_{i,1},\dots,d_{i,n_i})$, where $d_{i,j}$ is  the file index of  $j$'th request of user $i$. This is an i.i.d. sequence with the distribution $\mathcal{P}$.
\item
$\mathbf{q}(t) = (q_1(t),\ldots,q_L(t))$, where $q_i(t)$ is the queue length of Server $i$ at time $t$.
\end{itemize}

As indicated above, upon each request arrival at time $t$ an online mapping algorithm maps the request from the requesting user to a server with assuming full knowledge of servers' cache contents, servers' queue status $\mathbf{q}(t)$, and the communication cost matrix $\mathbf{C}$. This mapping can be formally defined as follows.
\begin{definition}
The mapping scheme of the above system is a function $\Psi$ defined as
\begin{equation}
\Psi \big( (i,j), \mathbf{C}, \mathbf{q}(t_{i,j}) \big) \mapsto \{1,\ldots,L\},
\end{equation}
where $(i,j)$ denotes the $j$th request of User $i$ and $t_{i,j}$ shows the arrival time of such a request.
\end{definition}
Moreover, we define
\begin{itemize}
\item
$\mathbf{c}_i=\left(c_{i,S_{i,1}},\dots,c_{i,S_{i,n_i}}\right)$, where $c_{i,S_{i,j}}$ denotes the communication cost of serving $j$'th request of User $i$. Here, $S_{i,j}$ is the server responsible for $j$th request of User $i$, which is determined by the mapping scheme as $S_{i,j} \triangleq \Psi \big( (i,j), \mathbf{C}, \mathbf{q}(t_{i,j}) \big)$.
\item
$\boldsymbol{\tau}_i=(\tau_{i,1},\dots,\tau_{i,n_i})$, where $\tau_{i,j}$ is the time interval from appending $j$'th request of User $i$ to $S_{i,j}$'s queue, until the request is served. 
\end{itemize}

Then for evaluating the performance of each specific mapping scheme $\Psi$, we consider two metrics. Our first metric is the \emph{average cost} per file delivery
  \begin{equation}
  \bar{C} (\Psi) \triangleq  \lim_{T \rightarrow \infty}\mathbb{E} \left[ \frac{1}{\sum_{i=1}^{K}{n_i}}\sum_{i=1}^{K}{\sum_{j=1}^{n_i}{c_{i,S_{i,j}}}} \right].
  \end{equation} 
Note that the dependency of $\bar{C}$ to the forwarding scheme $\Psi$ is through the random variables $S_{i,j}$.
  
Our second metric is the \emph{average waiting time} for each request, defined as follows
\begin{equation}
  \bar{D} (\Psi) \triangleq  \lim_{T \rightarrow \infty} \mathbb{E} \left[ \frac{1}{\sum_{i=1}^{K}{n_i}}\sum_{i=1}^{K}{\sum_{j=1}^{n_i}{\tau_{i,j}}} \right].
\end{equation}
Here $\bar{D}$ depends on the forwarding scheme $\Psi$ through the length of servers' queues.

In the above definitions, the expectations are with respect to the randomness of servers service times, users arrival requests process, and any other randomness introduced by a random mapping rule.

\begin{remark}
The model introduced above is general-enough to cover various content delivery scenarios. As an example, the servers in the model can represent the \emph{edge servers} of a CDN (say Akamai), which deliver content to costumers \cite{Akamai_Nygren_2010}.  As another example, the servers and users can be the same devices in a device-to-device (D2D) setup, where each device has cached some files, and requests some other files \cite{Mingyue_IT16}.
\end{remark}
    
\section{Load Balancing vs. Communication Cost}\label{sec:LoadBal_CommCost}

As already mentioned in \S\ref{sec:Introduction}, there exists a fundamental trade-off between average cost $\bar{C}$ and waiting time $\bar{D}$ in a distributed caching scenario. That is because when you look for low communication cost, upon each request arrival, you do not have much options from the servers pool to assign the request to. This will limit the load balancing power of the mapping algorithm. The above tension can be managed by the mapping rule used, \ie, $\Psi$. Formally, this trade-off is captured by the following optimization problem
\begin{equation}\label{eq:CostDelay_OptimizationProblem}
\begin{array}{rl}
& \min\limits_{\Psi} \alpha \bar{C} (\Psi) + (1-\alpha)  \bar{D} (\Psi)\\
\mathrm{s.t.} & \\
& d_{i,j} \in \mc{Z}_{\Psi \big( (i,j), \mathbf{C}, \mathbf{q}(t_{i,j}) \big) },\ \forall i\in [1:K], j\in [1:n_i],
\end{array}
\end{equation} 
for arbitrary value of $\alpha \in [0,1]$. In the optimization problem \eqref{eq:CostDelay_OptimizationProblem}, the minimization is over all mapping rules which have full knowledge of communication cost matrix $\mathbf{C}$, and instantaneous queue length status of all servers $\mathbf{q}(t)$. The parameter $\alpha$ determines the importance of each performance metric. Also the constraint ensures that the assigned server has cached the file in the cache content placement phase.

The solution to this optimization problem leads to the pareto-optimal trade-off curve between communication cost and waiting time. It should be noticed that the problem is hard to solve in the general case. Thus, in the following subsections, we propose three different algorithms as practical heuristic solutions to this optimization problem.

\subsection{Scheme I: Probabilistic Scheme Switching}
In order to gain more insight into the optimization problem  \eqref{eq:CostDelay_OptimizationProblem}, let us consider two special cases of $\alpha=0$ and $\alpha=1$.
Setting $\alpha=0$ in \eqref{eq:CostDelay_OptimizationProblem} puts the focus on minimizing the average waiting time. In particular, it can easily be observed that the optimal forwarding scheme $\Psi$ in this case is
\[
\Psi_1 \big( (i,j), \mathbf{C}, \mathbf{q}(t_{i,j}) \big) = \underset{k:\ d(i,j)\in\mc{Z}_k}{\arg\min} q_k(t_{i,j})
\]
which means assigning each request to the server with minimum load that has cached the file, without considering communication costs (see for example \cite{Opt_of_Short_Line_Discipline_JAP77}).

In contrast, for $\alpha=1$ the focus is on minimizing average communication cost. In this case the optimal mapping scheme is
\[
\Psi_2 \big( (i,j), \mathbf{C}, \mathbf{q}(t_{i,j}) \big) = \underset{k:\ d(i,j)\in\mc{Z}_k}{\arg\min} c_{i,k}
\]
that means upon arrival of each request, it will be forwarded to the server with minimum cost, without considering current load of servers.

The above two special cases focus on minimum waiting time and minimum communication cost, respectively. A natural probabilistic generalization which considers both metrics is presented in Algorithm~\ref{alg:Alg1}.

\begin{algorithm}[H]
\caption{Probabilistic Scheme Switching (PSS)} \label{alg:Alg1}
\begin{algorithmic}[1]
  \Require $(i,j)$, $\zeta$, $\mathbf{C}$, $\mathbf{q}(t_{i,j})$
  \State $\Lambda \leftarrow \left\{k | k\in[1:L], d_{i,j}\in\mc{Z}_k \right\}$
  \State Generate $x \in [0,1]$ uniformly at random
  \If{$x \leq \zeta$} 
  	\State Query $\{q_k(t_{i,j})\}_{k \in \Lambda}$ 
  	\State Assign the $(i,j)$ request to $\Psi_1 = \underset{k \in \Lambda}{\arg\min}\ q_k(t_{i,j})$
  \Else	
  	\State Assign the $(i,j)$ request to $\Psi_2= \underset{k \in \Lambda}{\arg\min}\  c_{i,k}$
  \EndIf	
\end{algorithmic}
\end{algorithm}

In Algorithm~\ref{alg:Alg1}, for each request, with probability $\zeta$ the minimum delay approach is used, and with probability $1-\zeta$ the minimum cost approach is used. This scheme mimics the situation where some requests have more stringent delivery delay requirements. Here, the ratio of such packets to total requests is determined by $\zeta$.  It should be noted that this algorithm needs on average $LM\zeta /N$ queue length queries for each assignment.

\subsection{Scheme II: Weighted Metrics Combination}
As discussed above the solution to the optimization problem in \eqref{eq:CostDelay_OptimizationProblem} for $\alpha=0$ and $\alpha=1$ can be obtained in a straightforward manner. However, if interested in both low communication cost and acceptable delay at the same time, we should consider other values of $\alpha$ as well. This naturally leads to the mapping scheme presented in Algorithm~\ref{alg:Alg2}. 

\begin{algorithm}[H]
\caption{Weighted Metrics Combination (WMC)} \label{alg:Alg2}
\begin{algorithmic}[1]
  \Require $(i,j)$, $\alpha$, $\mathbf{C}$, $\mathbf{q}(t_{i,j})$
  \State $\Lambda \leftarrow \left\{k | k\in[1:L], d_{i,j}\in\mc{Z}_k \right\}$ 
  \State Query $\{q_k(t_{i,j})\}_{k \in \Lambda}$
  \State $\beta_1 \leftarrow \sum_{k \in \Lambda} c_{i,k}$
  \State $\beta_2 \leftarrow \sum_{k \in \Lambda} q_k(t_{i,j})$
  \ForAll{server $k \in \Lambda$} 
  	\State $\eta(i) \leftarrow \alpha \frac{c_{i,k}}{\beta_1} +(1-\alpha)\frac{q_k(t_{i,j})}{\beta_2}$
  \EndFor	
  \State Assign the request to $\underset{k \in \Lambda}{\arg\min} \ \eta(k)$
\end{algorithmic}
\end{algorithm}

In summary, in Algorithm~\ref{alg:Alg2}, we assign the request to the server with minimum value $\eta(k)$ (called desirability value of user $k$), among those servers which have cached the request. For server $k$, the value of $\eta(k)$, is a weighted sum of communication cost and the current load of the server. The weight is determined by the parameter $\alpha$. It should be noted that this algorithm requires on average $LM/N$ queries for each assignment.

\subsection{Scheme III: The Power of Multiple Choices}
In Algorithm~\ref{alg:Alg3}, we propose a scheme that reveals the fundamental trade-off between these two objectives via a different approach motivated by the \emph{power of multiple choices} in the balanced allocations literature \cite{ABKU99} (also, see \cite{Mitzen_sur01}).

\begin{algorithm}[H]
\caption{Multiple Choices Scheme (MCS)} \label{alg:Alg3}
\begin{algorithmic}[1]
  \Require $(i,j)$, $\Delta$, $\mathbf{C}$, $\mathbf{q}(t_{i,j})$
  \State $\Lambda \leftarrow \left\{k | k\in[1:L], d_{i,j}\in\mc{Z}_k \right\}$ 
  \State Sort $\{c_{i,k}\}_{k\in\Lambda}$ and choose indices of the least $\Delta$ values: $k_1\le \cdots \le k_\Delta$
  \State Query $\{q_k(t_{i,j})\}_{k\in\{k_1,\ldots,k_\Delta\}}$ 
  \State Assign the $(i,j)$ request to $\Psi=\underset{k\in\{k_1,\ldots,k_\Delta\}}{\arg\min} q_k(t_{i,j})$ 
\end{algorithmic}
\end{algorithm}

In Algorithm 3, upon arrival of each request the corresponding user first determines the set of servers which have cached the requested file \ie, $\Lambda$. By sorting these servers based on their communication costs to this user, $\Delta$ servers with lowest communication costs are determined. Then, this user compares the queue length of these $\Delta$ servers, and assigns the request to the least loaded one. Therefore, this algorithm needs $\Delta$ queries for each assignment.

Algorithm 3 is motivated by the \emph{Supermarket Model} investigated  by Mitzenmacher et al. in \cite{Mitzenmacher01}. In this model we have a number of servers serving a request arrival process. If each request is randomly assigned to a server, the system will face large queue delays. However, suppose each request first randomly chooses $\Delta\ge 2$ servers, and is then assigned to the least loaded one. It is shown in \cite{Mitzenmacher01} that this will result in an exponential improvement in the average queue length of the servers.

\begin{figure}
\begin{center}
\includegraphics[width=0.5\textwidth]{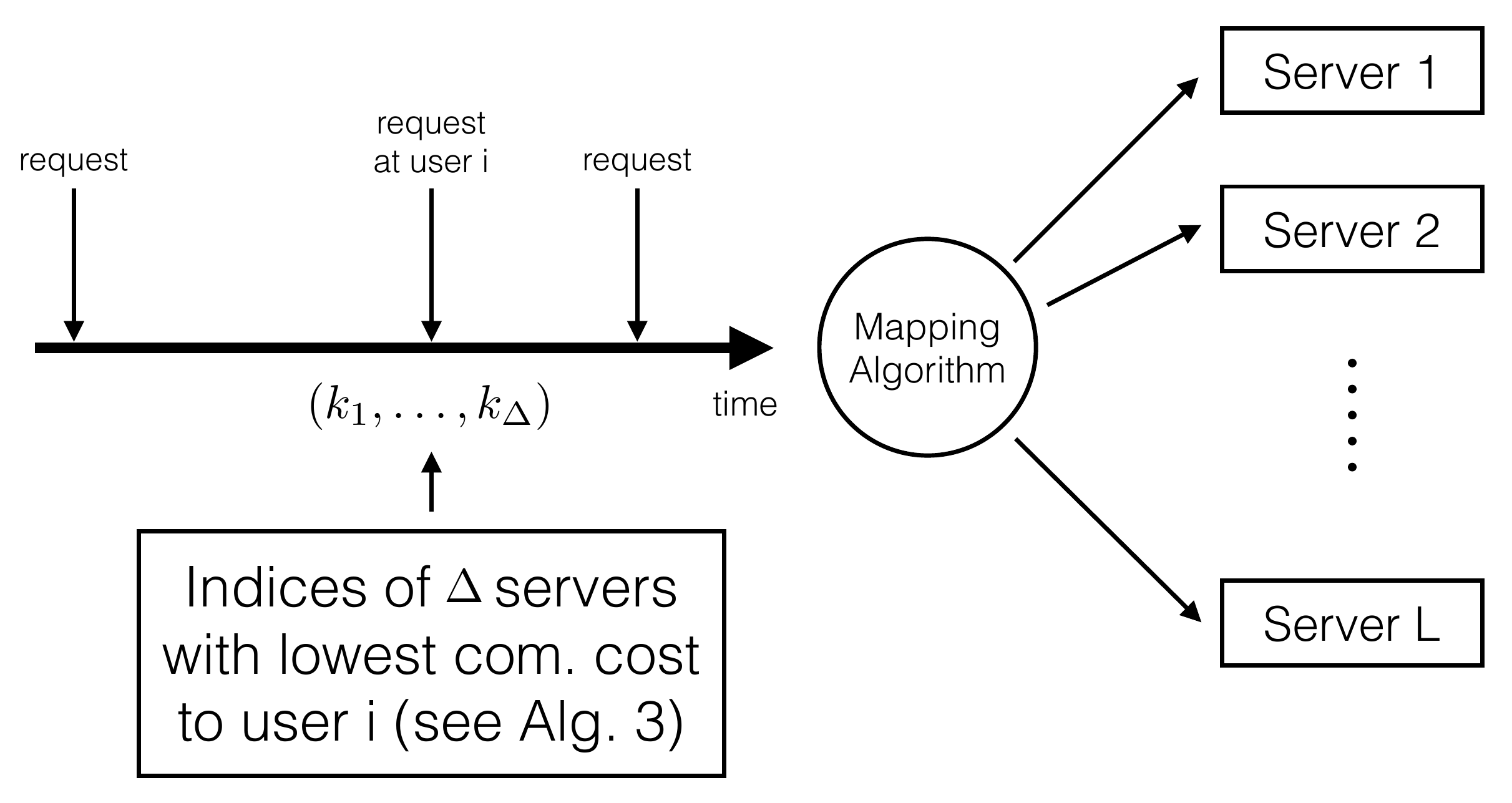}
\end{center}
\caption{Relation between Scheme III and the Supermarket Model studied  in \cite{Mitzenmacher01}.\label{Fig_SuperMarket}}
\end{figure}

In Algorithm 3, the metaphor of the results in \cite{Mitzenmacher01} is used to reduce the queuing delay, at the expense of higher communication cost (see Fig.~\ref{Fig_SuperMarket}). Here, we have $L$ servers serving requests from $K$ users. Since each request arrives at a distinct user every request will choose $\Delta$ random servers with low communication cost, based on its requesting user cost vector. Thus, a vector containing indexes of $\Delta$ candidate servers can be considered to be attached to each request, where this vector depends on the index of the user, and the communication cost matrix $\mathbf{C}$. If we choose large values of $\Delta$, then we have more options as our candidate servers to assign the request, which in turn reduces delay. However, this will introduce more communication cost, since servers with higher communication cost are let into the candidate set. Thus, $\Delta$ is the parameter managing the trade-off between delay and communication cost. 

\section{Performance Evaluation}\label{sec:PerfEval}
In this section, we investigate the performance of schemes proposed in Section~\ref{sec:LoadBal_CommCost} by developing an event-based simulation environment.

In our simulations, presented in the following, we assume that $L = K = 100$ and $N = 70$. The average incoming request rate for the $i$th user is $\lambda_i = 0.9$ and each servers' queue has an $\text{Exp}(1)$ service process. 
Each simulation consists of $10^5$ request events and every point on each graph is an average of $100$ independent simulation runs.
To generate the cost matrix $\mathbf{C}$, we consider that users and servers are distributed over a square Lattice uniformly at random. Then the communication cost between each user and a server is their Manhattan distance.

\begin{figure}
\begin{center}
\includegraphics[width=0.48\textwidth]{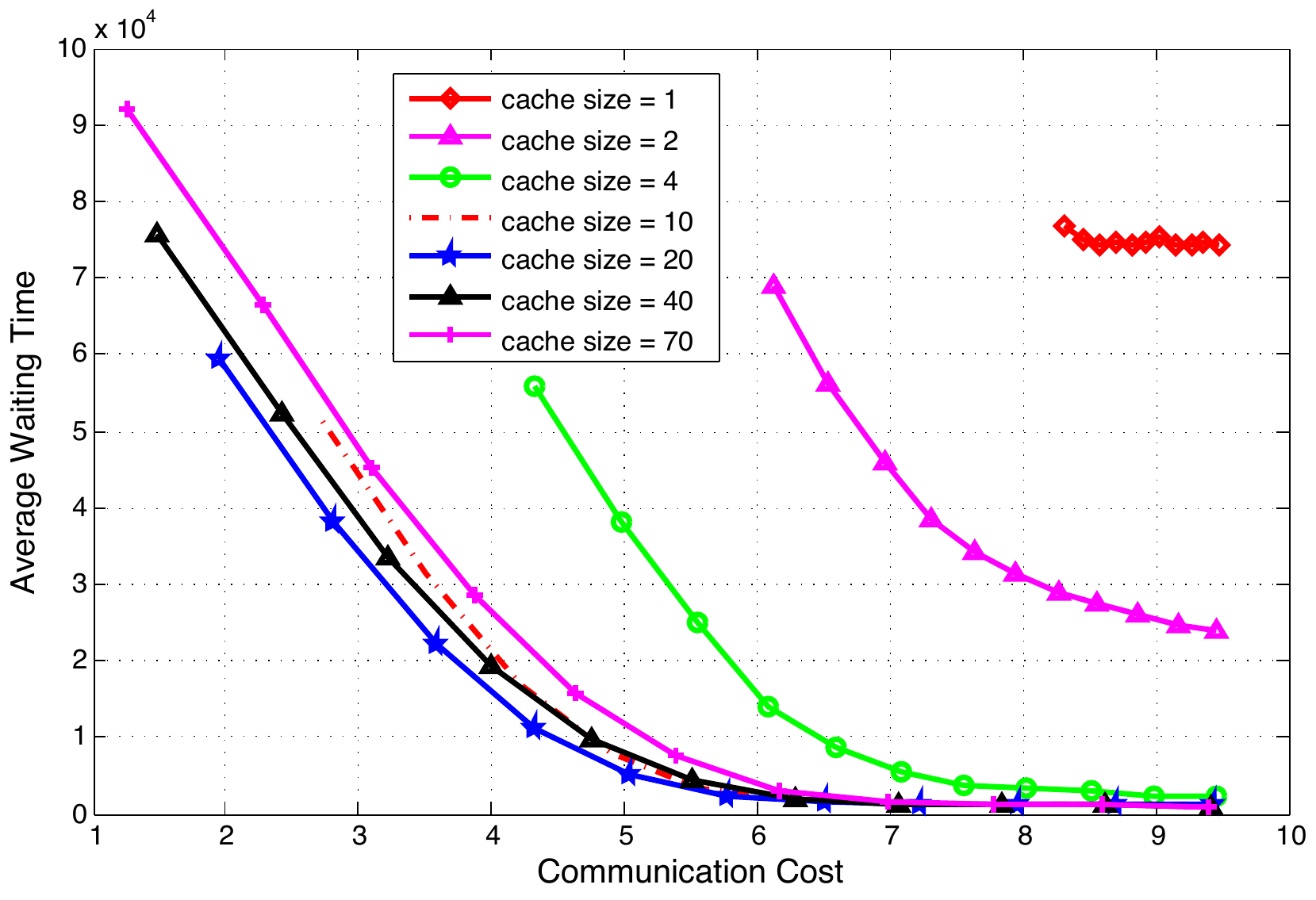}
\end{center}
\caption{Average waiting time, average cost trade-off for Algorithm~\ref{alg:Alg1} (PSS).}\label{fig:Alg1Performance}
\end{figure}

In the first set of simulations, the performance trade-off of Algorithm~\ref{alg:Alg1} is shown in Fig.~\ref{fig:Alg1Performance} for different cache sizes. For each curve on the figure, the most left point corresponds to $\zeta=0$, \ie, it leads to the minimum communication cost scheme. In contrast, the most right point corresponds to $\zeta=1$ which results to the minimum average waiting time scheme. As $\zeta$ is varied between $0$ and $1$ the trade-off curve of Algorithm~\ref{alg:Alg1} is captured. From the figure we observe that for low cache sizes, \ie, $M\le 4$, we have both high average waiting time and communication cost. In particular for $M=1$, the average waiting time cannot be reduced even by introducing large communication cost.

\begin{figure}
\begin{center}
\includegraphics[width=0.48\textwidth]{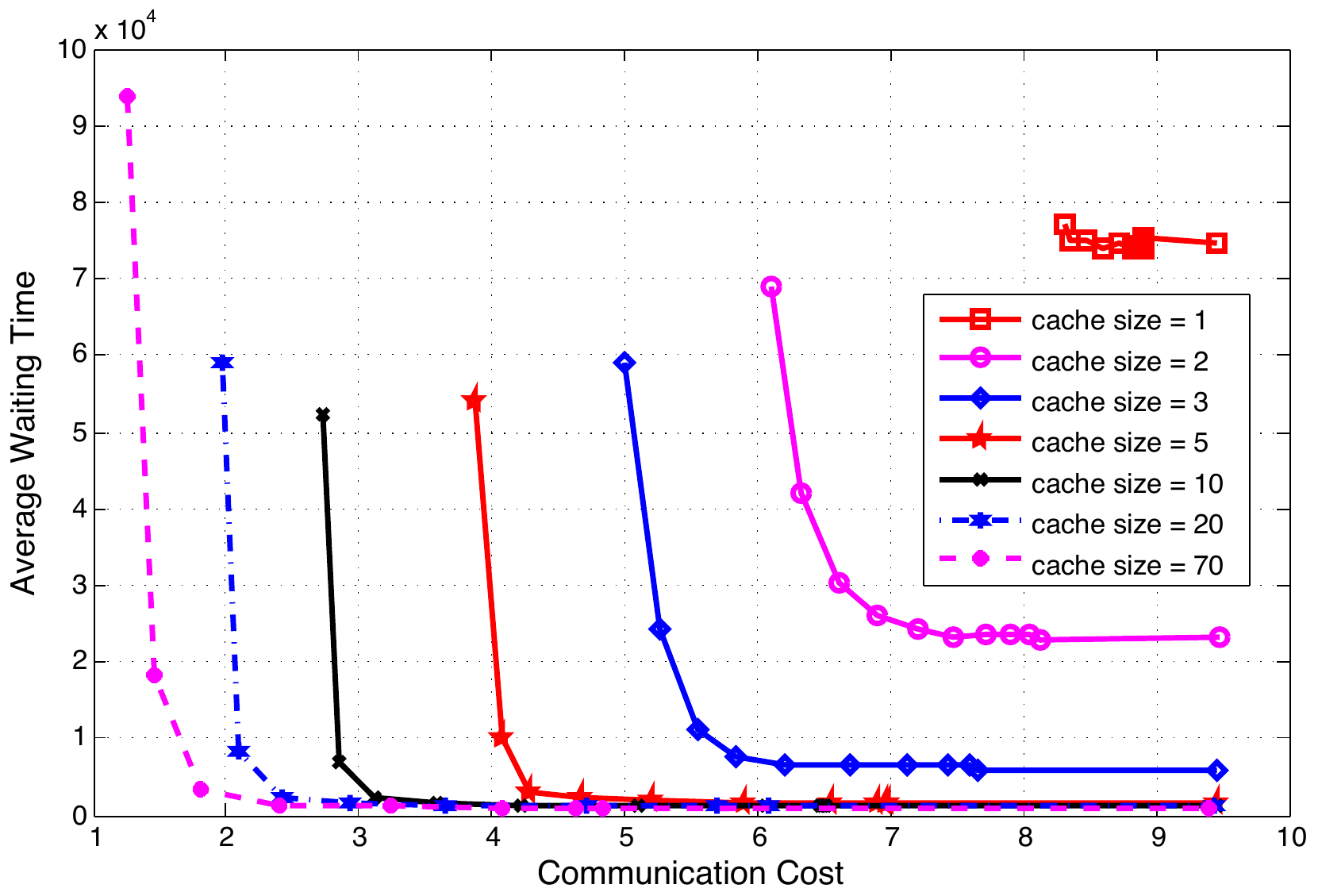}
\end{center}
\caption{Average waiting time, average cost trade-off for Algorithm~\ref{alg:Alg2} (WMC).}\label{fig:Alg2Performance}
\end{figure}

Fig.~\ref{fig:Alg2Performance} shows the performance trade-off of our second set of simulations. Here, the performance trade-off is captured by varying the parameter $\alpha \in [0,1]$ in Algorithm~\ref{alg:Alg2}. Interestingly, in contrast to Fig.~\ref{fig:Alg1Performance}, we can reduce the average waiting time by introducing just a small communication cost. Moreover, for the case $M=1$ we observe the same behaviour as of Fig.~\ref{alg:Alg1}.

\begin{figure}
\begin{center}
\includegraphics[width=0.48\textwidth]{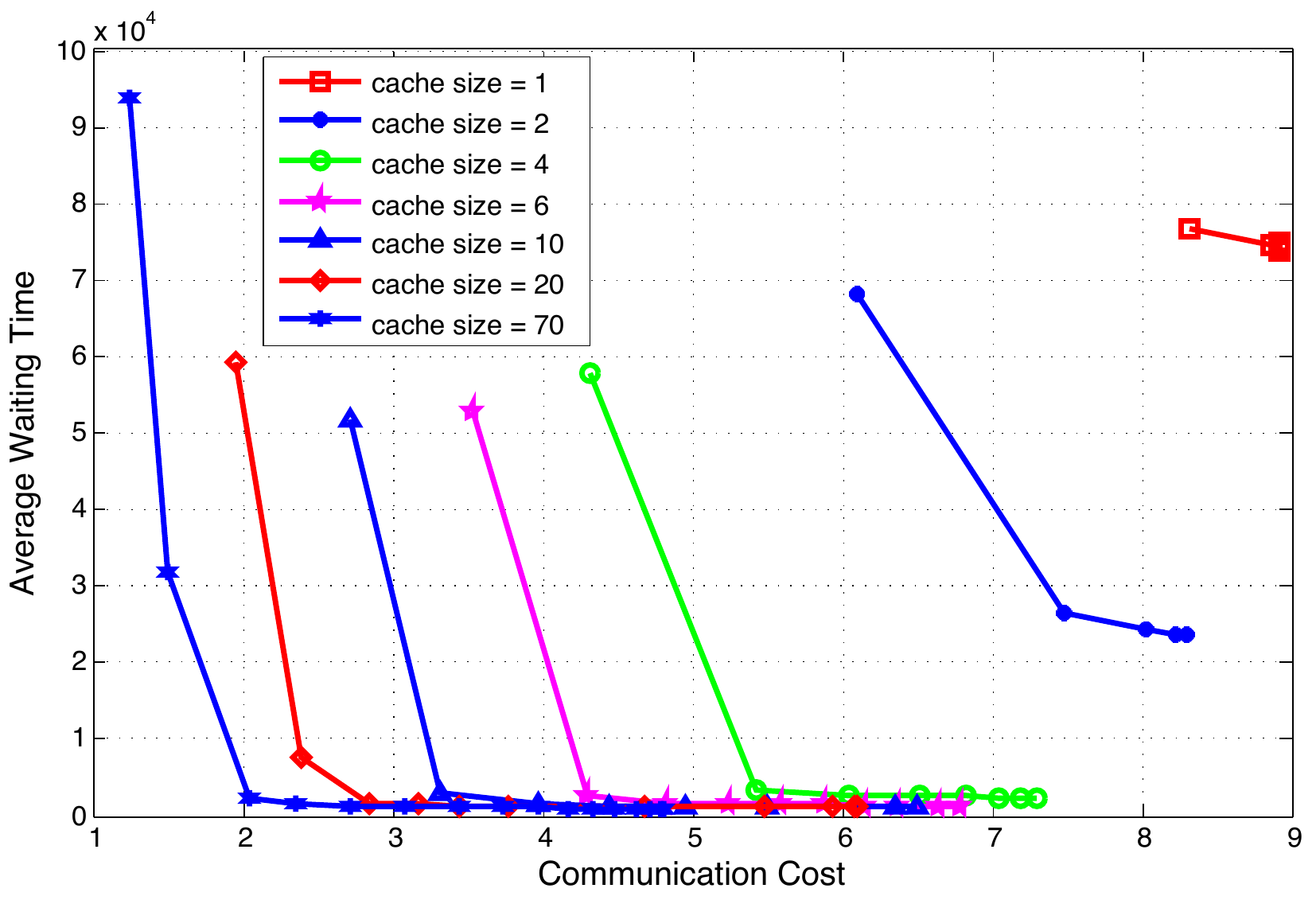}
\end{center}
\caption{Average waiting time, average cost trade-off for Algorithm~\ref{alg:Alg3} (MCS).}\label{fig:Alg3Performance}
\end{figure}

The third set of simulations, investigates the performance of Algorithm~\ref{alg:Alg3} shown in Fig.~\ref{fig:Alg3Performance}. Here the parameter $\Delta$ manages the trade-off between the average waiting time and communication cost. The performance of Algorithm~\ref{alg:Alg3} trade-off management is similar to that of Algorithm~\ref{alg:Alg2}; by introducing small communication cost the average waiting time is reduced significantly. Here, this phase transition can be explained by the well known power of two choices phenomenon discussed in \cite{ABKU99,Mitzenmacher01,Mitzen_sur01}.

\begin{figure}
\begin{center}
\includegraphics[width=0.48\textwidth]{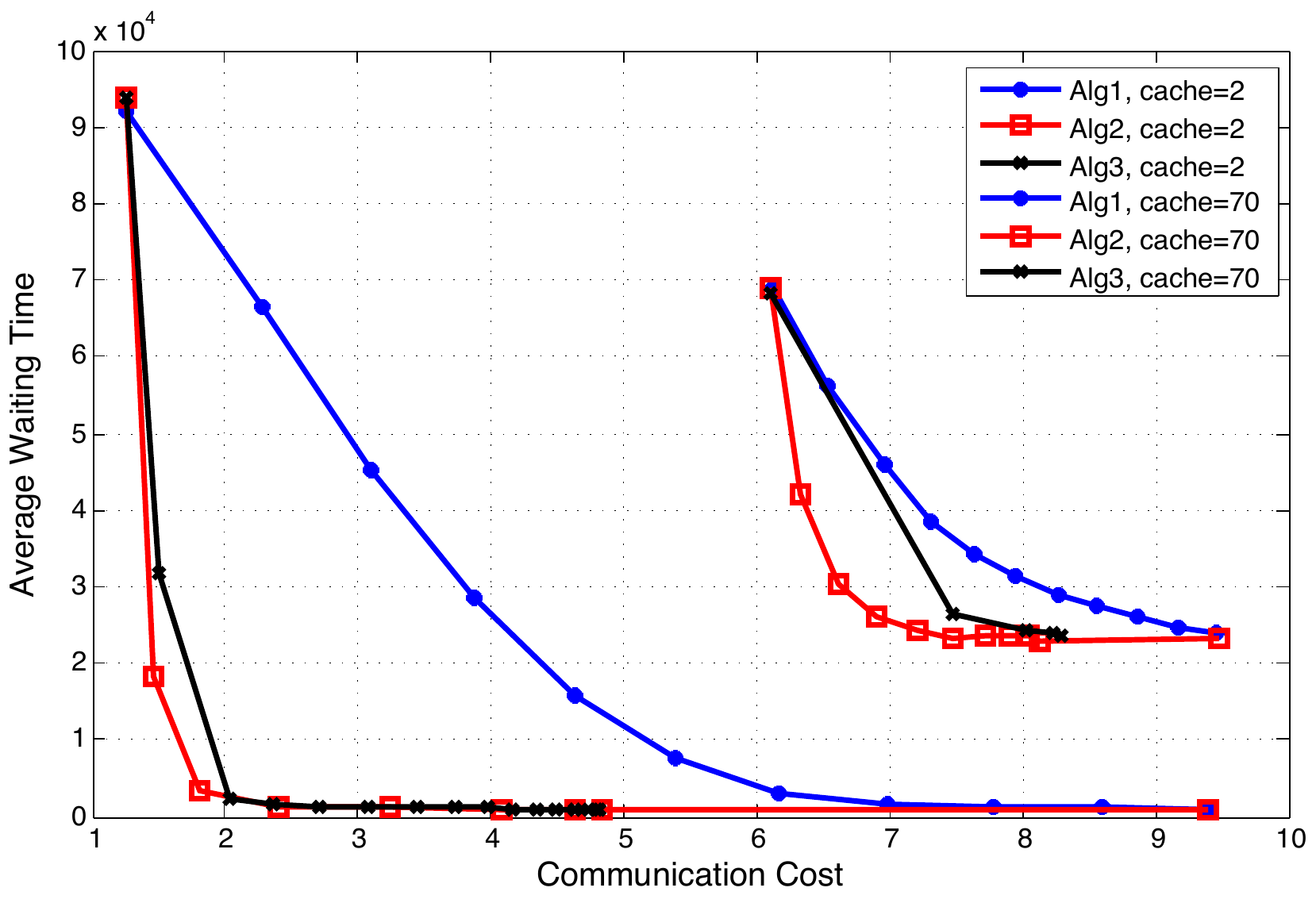}
\end{center}
\caption{Comparison of proposed algorithms for two different cache sizes.}\label{fig:PerformanceComparison}
\end{figure}

\begin{figure}
\begin{center}
\includegraphics[width=0.48\textwidth]{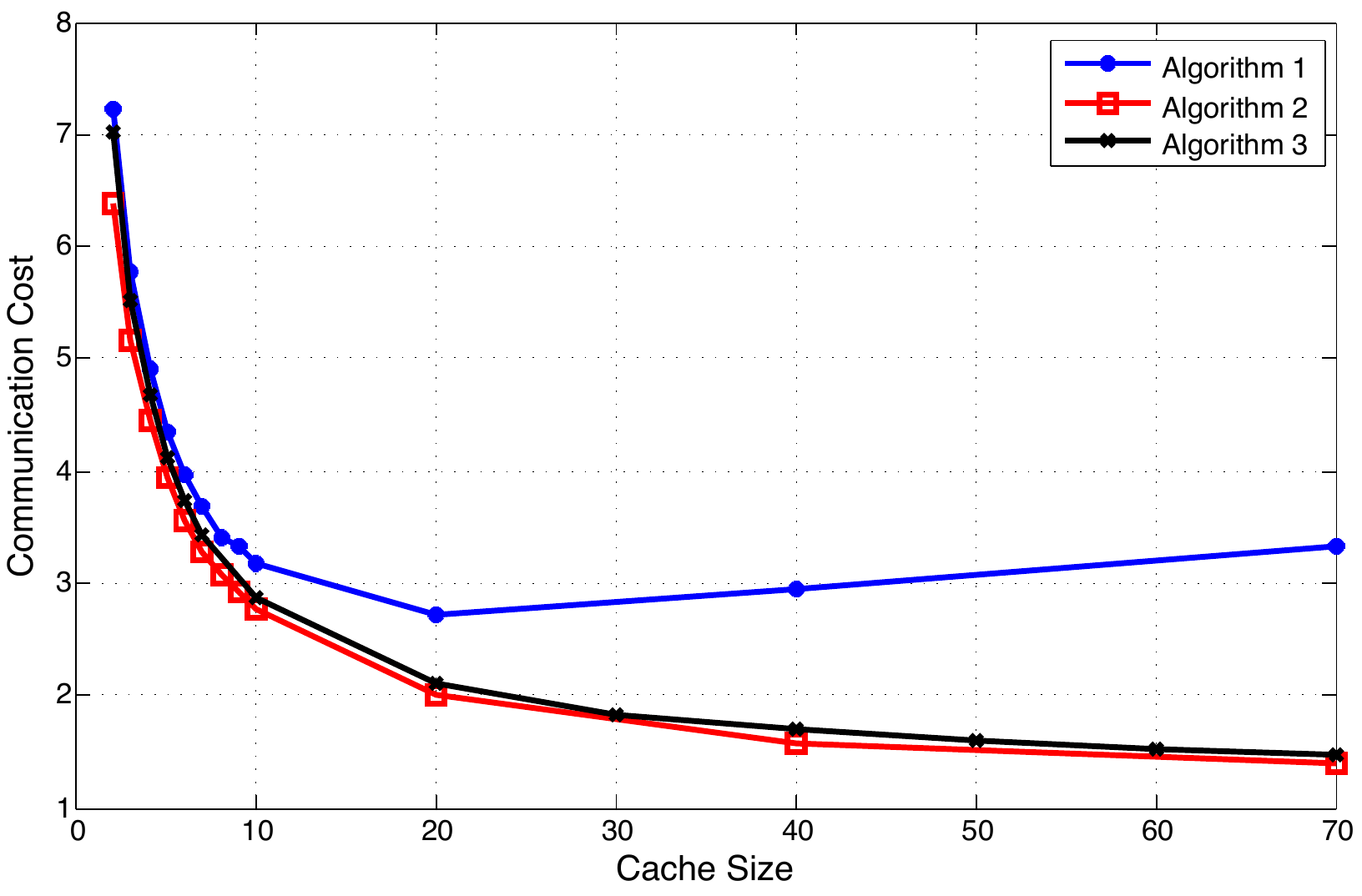}
\end{center}
\caption{Communication cost versus cache size when average waiting time is 40000 time slots.}\label{fig:cost_vs_cachesize}
\end{figure}

In order to compare the performance of the three proposed schemes, their trade-off curves are plotted together in Fig.~\ref{fig:PerformanceComparison} for two values of cache sizes, \ie, $M=2$ and $M=70$. In general, Algorithm~\ref{alg:Alg2} has a slightly better performance than Algorithm~\ref{alg:Alg3} and both of them surpass Algorithm~\ref{alg:Alg1}. It should be noted that, for large cache sizes, Algorithm~\ref{alg:Alg2} and \ref{alg:Alg3} have almost the same performance.
In addition in Fig.~\ref{fig:cost_vs_cachesize}, the communication cost versus cache size for three schemes is plotted for a fixed average waiting time. This figure confirms the trend observed above for Fig.~\ref{fig:PerformanceComparison}.

In summary, our simulation results show that Algorithm~\ref{alg:Alg2} has the best performance among all proposed schemes. In fact this algorithm solves an instantaneous optimization problem for each request assignment. Although this new optimization problem is different from our original problem \eqref{eq:CostDelay_OptimizationProblem} but they are closely related. This is why Algorithm~\ref{alg:Alg2} performs very well. However, the main drawback of this algorithm is   that upon each assignment, it needs to query all the nodes that have cached the requested file. Interestingly, Algorithm~\ref{alg:Alg3} addresses this drawback with negligible performance loss. This is due to power of two choices which is a well investigated phenomenon in load balancing literature. Finally, though Algorithm~\ref{alg:Alg1} has the worst performance, it has the capability of handling the requests with different delay requirements.
Table~\ref{tab:QueryCost} summarizes query costs of each algorithm.

\begin{table}
\begin{center}
{\renewcommand{\arraystretch}{1.5}
\begin{tabular}{ | c |c| }
  \hline
  Algorithm & Average Number of Queries Required \\
  \hline
  Algorithm~\ref{alg:Alg1} (PSS) & \mbox{$\zeta\frac{LM}{N}$} \\
  \hline
  Algorithm~\ref{alg:Alg2} (WMC) & $\frac{LM}{N}$ \\
  \hline
  Algorithm~\ref{alg:Alg3} (MCS) & $\Delta$ \\
  \hline
\end{tabular}}
\end{center}
\caption{Average query cost of each proposed mapping algorithm.}\label{tab:QueryCost}
\end{table}

\section{Conclusion}\label{sec:Conclusion}
We investigate the trade-off between the average waiting time and communication cost formally via a stochastic optimization problem introduced in \S\ref{sec:LoadBal_CommCost}. We show that although deriving the exact solution to the optimization problem is hard, our proposed heuristic solutions (\ie, mapping algorithms) can manage this trade-off in different scenarios. While our first algorithm is suitable for demands with different delivery delay requirements, second and third schemes achieve better trade-off curves.
Our results show that by sacrificing a small amount of communication cost one can arrive at a balanced network load.

\bibliographystyle{IEEEtran}
\bibliography{ICC17}

\end{document}